\documentclass[preprint,superscriptaddress,amsmath,amssymb,aps,pre]{revtex4-1}

\usepackage{graphicx}
\usepackage{dcolumn}
\usepackage{bm}
\usepackage{color}
\usepackage{bbold}
\usepackage{soul}
\usepackage{epsfig,epic}
\newcommand{\newc}{\newcommand}
\newc{\beq}{\begin{equation}}
\newc{\eeq}{\end{equation}}
\newc{\kt}{\rangle}
\newc{\br}{\langle}
\newc{\beqa}{\begin{eqnarray}}
\newc{\eeqa}{\end{eqnarray}}
\newc{\longra}{\longrightarrow}
\newc{\ovnu}{\overline{\nu}}
\newc{\ovk}{\overline{k}}

\newcommand{\jmin}{j_{\text{min}}}
\newcommand{\jmax}{j_{\text{max}}}
\newcommand{\tr}{\text{Tr}}
\newcommand{\Ut}{\tilde{U}}

\makeatletter
\let\Hy@backout\@gobble
\makeatother

\begin{document}

\title{Out-of-time-ordered correlator in the quantum bakers map and truncated unitary matrices}

\author{Arul Lakshminarayan}
\affiliation{Department of Physics, Indian Institute of Technology Madras, Chennai, India~600036}

\date{\today}

\begin{abstract}
The out-of-time-ordered correlator (OTOC) is a measure of quantum chaos that is being vigorously investigated.
Analytically accessible simple models that have long been studied in other contexts 
could provide insights into such measures. This paper investigates the 
OTOC in the quantum bakers map which is the quantum version of a simple and exactly 
solvable model of deterministic chaos that caricatures the action of kneading dough. Exact solutions based on 
the semiquantum approximation are derived that tracks very well the correlators till the Ehrenfest time. The growth occurs at the exponential rate of the classical Lyapunov exponent, but modulated
by slowly changing coefficients. Beyond this time saturation occurs as a value close to that of random matrices. Using 
projectors for observables naturally leads to truncations of the unitary time-$t$ propagator and the growth of their singular values
is shown to be intimately related to the growth of the out-of-time-ordered correlators.

\end{abstract}


\maketitle

\section{Introduction}

Quantum mechanics of low-dimensional systems with a deterministically chaotic classical limit  have attracted steady attention since the late 1970s, and textbooks such as \cite{Gutzwiller90, Haake, reichl2013transition, ozorio1989} chronicle a variety of these studies. These pose significant challenges and have given us various insights, including semiclassical periodic orbit theory and the relevance of random matrix ensembles for even one-particle systems whose classical limit is chaotic and surveys collected in \cite{GVZ1989} form an excellent introduction. 
A resurgence of interest in quantum chaotic or nonintegrable systems has occurred around the related themes of scrambling and out-of-time-ordered correlators or OTOC \cite{Rozenbaum17,Hashimoto2017,Jalabert2018,GarciaMata2018,ChenZhou2018,COTLER2018,Seshadri2018,Sondhi2018}. The OTOC, as commonly defined, is connected to the development of non-commutativity of initially commuting operators \cite{Maldacena2016} and therefore this forms a convenient starting point. In the context of many-body systems they 
capture how initially localized information spreads and is related to
the Lieb-Robinson bound for commutator growth in systems with a finite range of interactions \cite{Lashkari2013,Roberts2016}. 

In the recent past, it has been used in the study of quantum field theories and black holes, which are said to be nature's
fastest scramblers as they saturate a conjectured upper-bound on chaos \cite{Shenker2014,Maldacena2016,Roberts2016}. A, by now standard, qualitative motivation for relating commutators with chaos is that $-\br [\hat{q}(t),\hat{q}(0)]^2\kt \sim \hbar^2 \{q(t), q(0)\}^2=\hbar^2 (\partial q(t)/\partial p(0))^2 \sim \hbar^2 e^{2 \lambda t}$, where the semiclassical connection with Poisson brackets is used and further in the last step a chaotic evolution with a Lyapunov exponent of $\lambda$ is assumed. 
Thus the growth of the commutator is a quantum measure of instability, and the Lyapunov exponent growth is expected in a time regime that is between a diffusion time scale $t_d$ and the Ehrenfest time scale $t_{EF}$ at which quantum-classical correspondence, if any, breaks down.  The growth of the commutator being a purely quantum measure, can be used in systems such as spin chains which have no apparent classical limits and is a dynamical measure of the systems complexity. 

Simple models from classical dynamical systems have played an important role in the study of quantum chaos. These include quantum maps \cite{BBTV1979,IZRAILEV1990}, which are Floquet systems or quantizations of finite canonical transformations which 
have also been invoked in recent studies of OTOC. For example, the standard map or kicked rotor has been studied in \cite{Rozenbaum17} while the quantum cat map and its perturbed versions have been studied in the context of operator spreading and OTOC \cite{Sondhi2018,GarciaMata2018}. The classical cat map, introduced by Arnold and Avez \cite{ArnoldAvez}, is a smooth linear area-preserving map of the two-torus into itself. It's quantization \cite{BerryHannay} possesses nongeneric features such as exact 
periodicity in time, which is overcome by smooth perturbations on it but renders it analytically intractable. The classical bakers map, introduced by E. Hopf \cite{HopfBaker} is a discontinuous linear transformation of the phase space,
in the form of a square, into itself that is a caricature of the actions involved in kneading dough that leads to a uniform mixture, an essential prerequisite for good pastry. 
It is an exactly solvable and deterministic model of chaos \cite{Strogatz,Ott2002}, and yet is strongly stochastic in the sense that it is isomorphic to a Bernoulli process, in other words it is as random as a coin toss \cite{ORNSTEIN1989} and has been described as the ``harmonic oscillator of chaos". 

The quantization of the bakers map, which treats the phase space square as a torus, has been studied in many different flavors since the original quantization by Balazs and Voros \cite{BalazsVoros1987,BalazsVoros1989}. The quantum map consists of discrete Fourier transforms on appropriate spaces and is hence a simple $N-$dimensional  unitary matrix $B$ which has also been implemented in an NMR experiment \cite{NMRBaker1,NMRBaker2}. It can be considered, when the Hilbert space dimension $N=2^L$, as dynamics of $L$ qubits with nonlocal interactions and entanglement in these qubits have also been studied \cite{ScottCaves2003}. However, despite its simplicity, it has not yet been solved analytically, say for its eigenspectra which have a close resemblance to those of random matrix ensembles \cite{MLMehta}. In this sense it is arguably less valuable than its classical counterpart. Nevertheless there are some simplifications in the sense that the powers of the classical map can be independently quantized and a simpler operator results that is not the powers of the quantum map itself \cite{SaracenoVoros1994,ArulNLB1994nuclear}. This ``delayed quantization" has been called
semiquantum and provides a valuable approximation $B_t$ for the time evolved propagator (the powers of the matrix $B^t$). For example this is the starting point for a semiclassical periodic orbit quantization of the bakers map \cite{AlmSara1991}.

The semiquantum approximation is used below to evaluate analytically the OTOC for the quantum bakers map. By contruction the semiquantum approximation is valid only till the Ehrenfest time and therefore the OTOC can be explicitly found in its Lyapunov regime. The analytical results reveal a surprisingly complex situation with the rate reaching the classical Lyapunov exponent, $\lambda$, (and {\it not twice} this) at late times. The origin for this could be from the dynamics being non-smooth, giving rise to ``diffraction" effects. It is known \cite{DegliEsposti2006} that sufficiently localized operators are needed for quantum classical correspondence to exist, localized enough to not suffer the discontinuities within the Ehrenfest time. The other is also that the operators themselves need not have smooth classical symbols with which to compute their Poisson or Moyal brackets. For the bakers map it is the former property that results in this rate being different from $2 \lambda$ and closer to $\lambda$, as we see this also for operators with smooth classical limits. However in this work, we consider for analytical purposes phase space projection operators whose classical limits are evidently characteristic functions on phase space. This allows for the OTOC to be
written exclusively based on a truncation of the unitary propagator $B^t$ to a non-unitary operator. That the spectrum and singular values of such truncations carry information about scrambling is a general feature.

Consider the non-commutativity of an operators $A(0)$ and $A(t)$ as given by 
\beq
\begin{split}
f(t)&=-\frac{1}{2}\text{Tr} [A(t),A(0)]^2=f_2(t)-f_4(t),\\
f_2(t)&=\text{Tr}(A(t)^2 A(0)^2),\;f_4(t)= \text{Tr}(A(t) A(0) A(t) A(0)),
\end{split}
\eeq
where $A(t)=U^{-t} A(0) U^{t}$ is the operator evolved to time $t$ by the dynamics of the propagator $U^t$. The term $f_2(t)$
is a two-point correlation, while $f_4(t)$ is a four-point OTOC. 
Let the operator $A(0)$ be a projector:
\beq
P(0)=\sum_{j=\jmin}^{\jmax} \; |j \kt \br j |, 
\label{eq:proj}
\eeq
where $\{|j\kt, \, 0 \le j \le N-1 \}$ forms a complete orthonormal basis. Let $J=[\jmin,\jmax]$ be the range of the projector, and let the complementary range be $\overline{J}=[0,\jmin-1] \cup [\jmax+1,N-1]$. We will also use the same letter $J$ to denote the dimensionality $\jmax-\jmin+1$ of the projector space. It follows that 
\beq
f_2(t)=\tr(P(t)P(0))=\tr(\Ut^{t\dagger}\Ut^t)=\|\Ut^t\|^2,
\label{eq:f2oft}
\eeq
where $P(0) U^t P(0)=\Ut^t$ is a $J-$dimensional truncation of $U^t$ in the basis $\{|j \kt \}$ and the norm is Euclidean.

Such correlations have been previously studied for a variety of ``interacting" bakers maps, that are isomorphic to 
Markov chains in \cite{LakBalJSP}. These quantized maps studied as models of relaxation in classical mixing systems \cite{Elskens1985}.
In general if $g$ and $h$ are two functions on a phase
space, then $\br h_t g \kt -\br h_t\kt \br g \kt$ (where $h_t$ is the time-evolved function and $\br \cdot \kt$ denotes phase space averaging), decay exponentially at the rates determined by the Ruelle-Pollicott resonances \cite{Ruelle1986}.
The change in the two-point correlator is rapid in comparison with the OTOC and is essentially governed in the classical limit by these resonances that lead to mixing. Beyond the time-scales set by these, the non-commutativity grows due to the decay of the OTOC. This is the process we are interested in, but we will study the non-commutativity $f(t)$ which includes both these contributions and sometimes loosely refer to it as OTOC itself.

The OTOC has the following simplifications:
\beq
f_4(t)=\sum_{j,j' \in J} \; |\br j |P(t)| j' \kt |^2=\tr(\Ut^{t\dagger}\Ut^t)^2=\|\Ut^{t \dagger} \Ut^t\|^2.
\eeq
Thus if the eigenvalues of $\Ut^{t\dagger}\Ut^t$, or equivalently (square of) the singular values of $\Ut^t$, are $\mu_i(t)$, then
these completely determine the $f_2(t)$ as well as $f_4(t)$. Their difference $f(t)$ is then 
\beq
f(t)=\sum_{j \in J} \sum_{j' \in \overline{J}} \; |\br j |P(t)| j' \kt |^2=\sum_{i=1}^{J}\mu_i(t)\left(1-\mu_i(t)\right).
\label{eq:foft}
\eeq
Truncated unitary matrices, especially from random matrix ensembles, have been studied vigorously since the pioneering work
of Zyczkowski and Sommers \cite{KarolSommers2000} and find applications in many contexts such as chaotic scattering, where truncations of $S$-matrices arise \cite{TruncUnitary2,TruncUnitary1}, open quantum systems \cite{TruncUnitary4}, tunneling studies \cite{TruncUnitary3}. Note that $0 \le \mu_i(t) \le 1$ and 
$f(t)\le J/4$. The growth in norm of truncations of powers of unitary matrices are naturally related to the OTOC, an observation that may 
provide more insights. Apart from the singular values $\mu_i(t)$ of $\Ut^t$, their complex eigenvalues also provide
a characterization of the dynamics that reflects the growth of the OTOC as will be seen in the example of the bakers map.

\section{Preliminaries of the bakers map}

This section is to provide an introduction (to non-bakers) of  well-known facts of the bakers 
map both classical and quantal.
\begin{figure}[h]
\includegraphics[scale=.3]{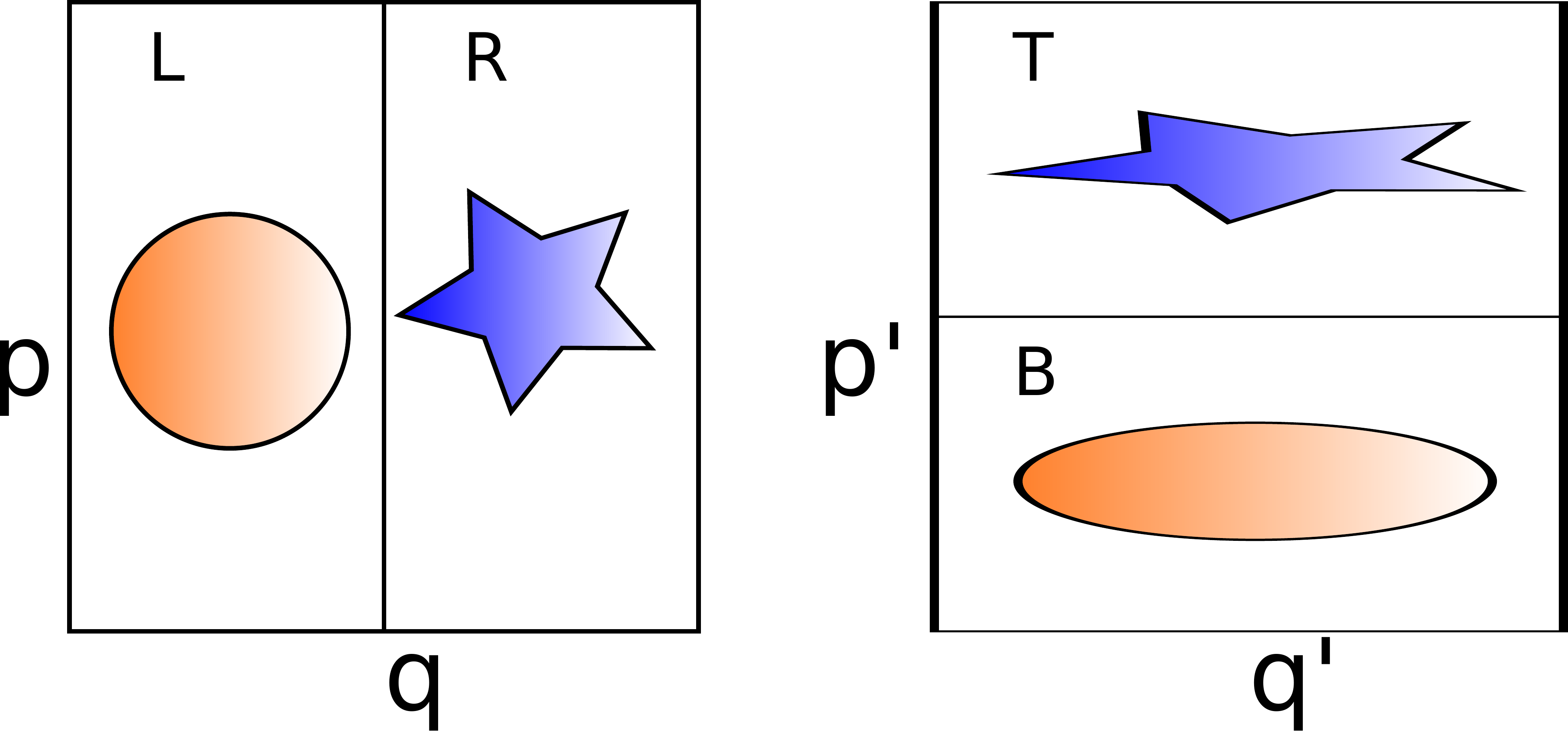}
\caption{The bakers map action on the unit square on the left takes it to the right, by stretching the left half $L$ by a
factor of $2$ along $q$ and compressing by a factor $1/2$ along $p$ so that it becomes the bottom half $B$. 
The transformation from the right half $R$ to the top half $T$ is similar. Repeating this action constitutes a highly efficient mixing protocol and a solvable textbook example of deterministic chaos. }
\label{fig:classbaker1}
\end{figure}
The classical bakers map is given by the transformation of 
$(q,p) \in [0,1) \times [0,1)$ to itself and is given by 
\beq
\mathcal{T}(q,p)=(q',p')=(2q \;\left(\text{mod}\; 1),(p+[2q])/2\right).
\eeq
It is piecewise linear with a  discontinuity at $q=1/2$, and 
if the square is treated as a torus it is discontinuous at $q=0$ as well.
The action on the unit square is illustrated in the Fig.~(\ref{fig:classbaker1}),
where the stretching by a factor of 2 along the $q$ direction and compression along the 
$p$ is illustrated. The left vertical half $L$ gets mapped into the bottom horizontal half 
$B$. In this action the $q$ suffers the ``doubling map" $q \mapsto 2q \; (\text{mod}\; 1)$ and the 
dynamics in terms of binary representation is one of the left-shift \cite{Strogatz,Ott2002}. The momentum ensures that the shifted bits
are not lost. If $q=0.a_0 a_1a _2 \cdots$ and $p=0.a_{-1}a_{-2} \cdots$ are the respective binary representations
($a_i \in \{0,1\}$), then $q'=0.a_1 a_2 a_3 \cdots$ and $p'=0.a_0a_{-1}a_{-2} \cdots$. Thus this left-shift iterated 
is the bakers map action that lays bare the heart of deterministic chaos. In particular the Lyapunov exponent is $\ln 2$,
all its orbits are hyperbolic (unstable), the map is ergodic and mixing.
The exponential growth of the number of periodic orbits is determined by the topological entropy which is also $\ln 2$.
The enumeration of periodic orbits and their structure plays a crucial role in the semiquantum operator as well.
Let
\beq 
\nu=\sum_{k=0}^{t-1}a_k 2^{k}, \;\; \text{and} \;\; \ovnu= \sum_{k=0}^{t-1}a_k 2^{t-k-1}
\label{eq:nunubar}
\eeq
be the binary expansion of an integer $\nu$ $(0 \le \nu \le 2^t-1)$ and $\ovnu$ is an integer whose binary expansion contains the corresponding 
bit-reversed string, read from right to left. The period$-t$ points are at 
\beq
q_{\nu}=\frac{\nu}{2^t-1},\;\; p_{\nu}=\frac{\ovnu}{2^t-1},
\eeq
and there are $2^t$ of these. Thus the classical map is in many ways exactly solvable, if fully chaotic, and moreover is
a caricature of what happens in the neighborhood of homoclinic intersections of stable and unstable manifolds
that are the genesis of Hamiltonian chaos \cite{Ott2002}.

The quantization is complicated by the lack of a Hamiltonian, even a time-dependent one, such as exists for the 
standard map or the kicked top, other well studied models of low-dimensional chaos. Nevertheless Balazs and Voros
observed that the generating function of the transformation from the left half $L$ to the bottom half
$B$ is $F_2(q,P)=2qP$. From the correspondence of the unitary propagator being the exponential of the 
classical generating function, the {\it mixed} representation of the transformation of $L\mapsto B$ is 
$\br P |B|q \kt \sim e^{-i 2 qP/\hbar}$. As the phase space is now a compact torus, the quantization is also 
one which takes this into account and subsequently the Hilbert space is a finite one, its dimension $N=A/h$, where
$A$ is the area of the torus, which we take as $1$. Hence $h=1/N$ is the effective Planck constant and the position states 
labeled by $|n\kt$, $0\le n \le N-1$ are related to the momentum states via the discrete Fourier transform. Thus the quantization 
of the bakers map was proposed to be the unitary operator, written in position basis to be:
\beq
B=G_{N}^{-1} \left(\begin{array}{cc} G_{N/2}&0\\0&G_{N/2}\end{array}\right),
\eeq
where
\beq
\br m |G_{N}|n \kt =\frac{1}{\sqrt{N}} \exp\left[ -\frac{2 \pi i}{N} (m +1/2)(n+1/2)\right].
\eeq
is the discrete Fourier transform. The shifts by $1/2$ were added by Saraceno  to restore parity symmetry,
which also enabled a detailed study of eigenstates and time-evolving coherent states in the quantum baker \cite{Saraceno1990}.
The $R \mapsto T$ part of the transformation is the lower block $G_{N/2}$. The factor of $1/2$ in the Fourier 
transforms originates from the stretching of the classical baker being by the factor $2$. Thus propagation of states by the bakers maps are
given by $B^t|\phi_0\kt$ and operators evolves as $B^{-t} A(0) B^t$, with $t$ being integers, but there are no simple forms for the
powers $B^t$, a fact related to the lack of analytical understanding of the spectra.

\begin{figure}[h]
\includegraphics[scale=.15]{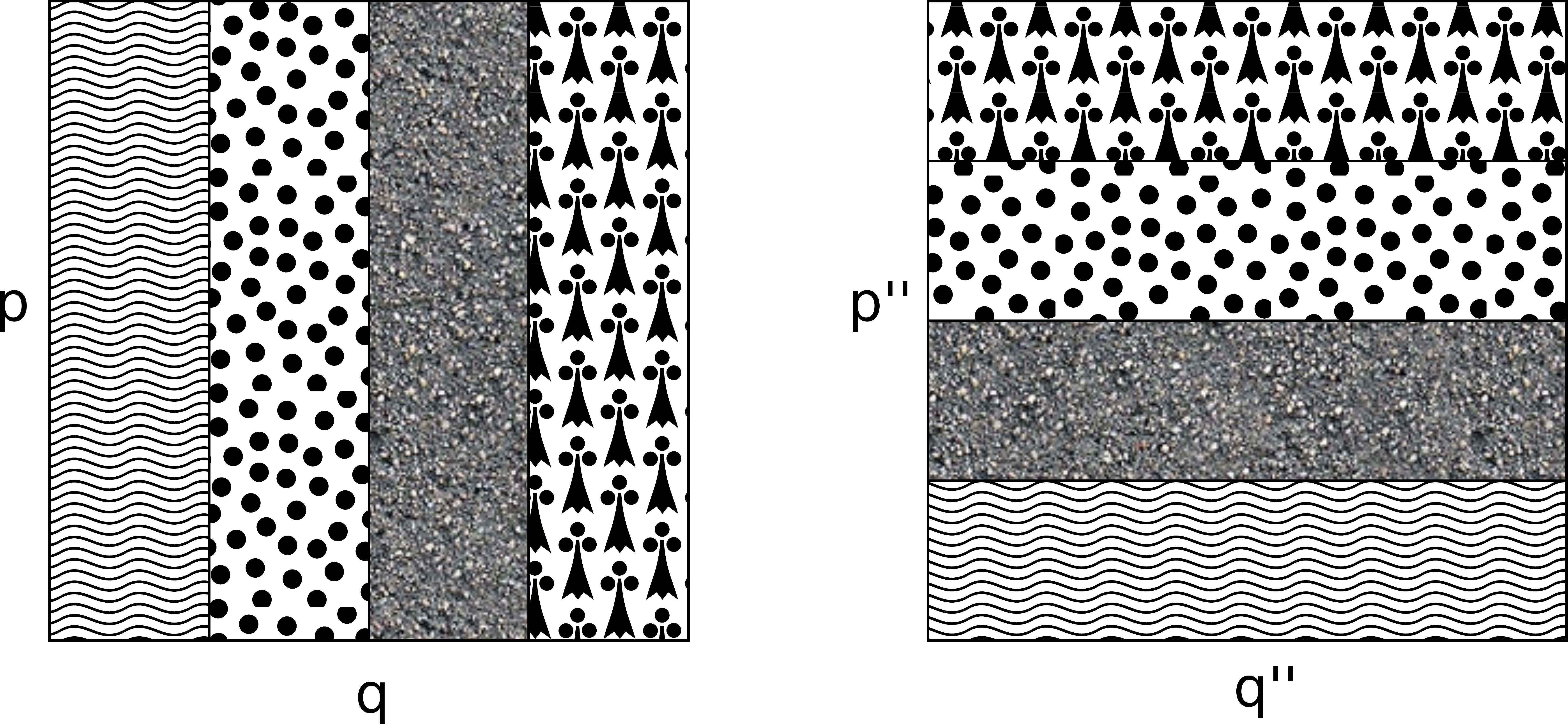}
\caption{Action on the unit-square phase space of the bakers map iterated twice. The $4$ identical vertical rectangles are each stretched and compressed by a factor of $4$ into corresponding horizontal rectangular partitions. Notice that unlike in Fig.~(\ref{fig:classbaker1}) the patterns in the partitions are not faithfully stretched in this illustration.} 
\label{fig:classbaker2}
\end{figure}

The time $t$ {\it semiquantum} propagator, $B_t$ is constructed by quantizing the classical baker iterated $t$ times \cite{SaracenoVoros1994}. This is unique to the baker map, and  $B^t \ne B_t$, but it is believed that $B_t \approx B^t$ till the time that $B_t$ can be defined which is at most the Ehrenfest time of $\log_2 N$.  When $N=N_0 2^{T}$, where $N_0$ is an odd integer, it is defined
till time $T$ and hence is the longest when $N$ is a power of $2$. 
For example $B_2$ is got on quantizing $\mathcal{T}^2$, whose action on the unit square is shown in Fig.~(\ref{fig:classbaker2}). This takes the 4 vertical partitions
of width $1/4$ to corresponding horizontal partitions of height $1/4$. Each of them can be quantized by the action of $G_{N/4}$ in the mixed representation. The position of the resulting 4 blocks is dictated by the period $2$ orbits which are the intersections of the vertical and corresponding horizontal partitions: ($00.00,01.01,10.10,11.11)$  In general the operator $B_t$ is governed by the $2^t$ fixed points of the classical time $t$ map (or equivalently the period$-t$ points of the classical map).
With these definitions:
\beq
B_t =G_{N}^{-1} (\mathcal{I}_t \otimes G_{N/2^t})
\eeq
where $\mathcal{I}_t$ is a $2^t \times 2^t$ matrix whose entries are zero except elements $(\mathcal{I}_t)_{\nu, \ovnu}=1$, where $\nu$ and $\ovnu$ are
given as in Eq.~(\ref{eq:nunubar}) which determine the classical periodic orbits. For the special case of $t=1$, $\mathcal{I}_1$ is the diagonal $2 \times 2$ identity matrix and one gets that $B_1 =B$.
At $t=2$, the $(\nu, \ovnu)$ pairs are $(0,0),(1,2),(2,1),(3,3)$ and $\mathcal{I}_2$ is a ``two-qubit swap gate",
hence
\beq
B_2=G_{N}^{-1} \left( \begin{array}{cccc} G_{N/4}&0&0&0\\0&0&G_{N/4}&0\\0&G_{N/4}&0&0\\0&0&0&G_{N/4}
\label{eq:semiqBt}
\end{array}\right).
\eeq
If $N=2^T$, $B_T$ will in the mixed representation (the second matrix)
will consist of $2^T$ $c-$ numbers, beyond that this is not defined, the classical partitions have reached the size of $\hbar$. This is also
the ``log-time" or the Ehrenfest time $t_{EF}=\ln(1/h)/\lambda =\log_2 N$, beyond which even initially maximally localized states suffer interference.

\section{Out-of-time-ordered correlator}

\begin{figure}[h]
\includegraphics[scale=1,angle=-90]{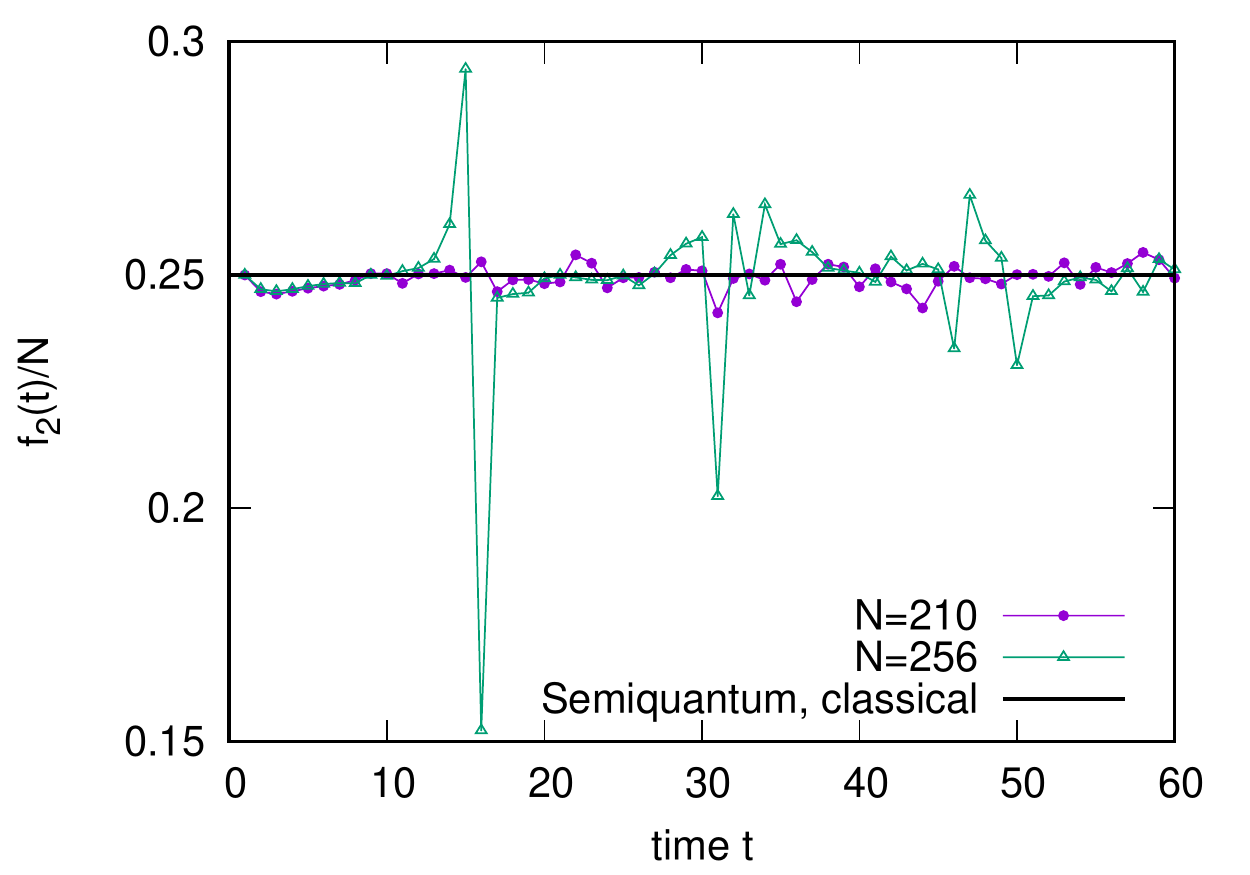}
\caption{ The two-time correlator $f_2(t)$ in Eq.~(\ref{eq:f2oft}) is shown as a function of time for two value of $N$. Both reach values close to $N/4$, while the case when $N$ is a power of $2$ shows anomalous oscillations after the log-time which is $\approx 8$. }
\label{fig:f2}
\end{figure}

The operator we choose is the projector $P(0)=\sum_{n=\jmin}^{\jmax}|n \kt \br n|$, where $|n \kt$ are position eigenstates, and which has a clear classical 
limit. If $\jmin=0$ and $\jmax=N/2-1$ it is the characteristic function of the left half vertical partition, the rectangle $L$ shown in Fig.~(\ref{fig:classbaker1}),  The quantity of interest is 
\beq
f(t)=-\frac{1}{2}[ P(0),P(t)]^2=\|\tilde{B}^t\|^2-\|\tilde{B}^{\dagger t} \tilde{B}^t \|^2,
\label{eq:Def_nof_F}
\eeq
where $\tilde{B}^t$ is the $J$ dimensional truncation of $B^t$.
Figure.~(\ref{fig:f2}) displays a normalized correlator $f_2(t)/N$ for the case when $\jmin=0$ and $\jmax=N/2-1$ for two cases of $N=210$ and $N=256$. The former case shows small fluctuations around $1/4$, while for the latter there are large fluctuations observed at $2 \log_2N$, twice the log-time and possibly multiples of these with decreasing amplitude. Note that $f_2(0)=N/2$, thus there is an instantaneous change to values close to $N/4$ for $t>0$.
This reflects the immediate mixing of the $L$ partition, in the sense that the classical characteristic function spreads so that $1/2$ of it is always in $L$ for all subsequent times. To control the rate of mixing, coupled bakers maps were studied with tunable coupling in \cite{Elskens1985} whose quantum versions were studied \cite{LakBalJSP}. See also \cite{Ostruszka2003} for such systems on a spherical phase space. The large fluctuations in $f_2(t)$ when $N$ is a power of $2$, especially at twice the log-time is consistent with known eccentricities of the quantum baker. This results in strongly localized eigenstates that are in fact 
all multifractal. Approximate subsets of eigenstates can be in this case constructed based on the ubiquitous, self-similar, binary Thue-Morse sequence 
and its generalizations \cite{MeenArulPRE}. However these eccentricities are prominent only at times beyond the log-time and therefore for the present purpose
they do not really concern us, for example the two cases of $N$ in Fig.~(\ref{fig:f2}) are essentially the same before the log-time.

While we cannot compute analytically $f_2(t)=\|\tilde{B}^t\|^2$ even for the bakers map above, the semiquantum $f_{2SQ}(t)=\|\tilde{B}_t\|^2$ turns out to be exactly $N/4$ for $t>0$ and hence is completely consistent with the classical. To begin we write $B_t$ in the mixed (momentum-position) basis
and denote momentum states as $|\tilde{m} \kt$. This is the matrix $\mathcal{I}_t \otimes G_{N/2^t}$, but it helps to write it explicitly as
\beq
\br \tilde{m}|B_t|n \kt = \sqrt{\frac{2^t}{N}} \exp\left[ \frac{-2^{t+1}\pi i}{N}\left( m+\frac{1}{2}-\frac{\nu N}{2^t}\right)\left( n+\frac{1}{2}-\frac{\ovnu N}{2^t}\right)\right] \Theta_{m\nu}\Theta_{n \ovnu}.
\label{eq:semiqmixedrep}
\eeq
For any $(m,n)$ pair ($m$ and $n$ take values in $[0, N-1]$) there exists a unique $\nu$ and hence $\ovnu$ as $\mathcal{I}_t$ is a permutation matrix. 
The connection is explicit in the function $\Theta_{m\nu}$  which is $=1$ if $\nu N/2^t \le m \le (\nu+1) N/2^t-1$ and $0$ otherwise. Multiplying by $G_N^{-1}$ on the left of the above mixed representation results in the matrix element of $B_t$ in the position basis:
\beq
\begin{split}
\br k |B_t|n \kt=&\frac{2^{t/2}}{N} e^{i \pi \ovnu} \exp\left[\frac{2\pi i \nu}{2^t}\left(k+\frac{1}{2}\right)\right]  \Theta_{n \ovnu}\\ & \times \sum_{m=0}^{N/2^t-1} \exp\left[ \frac{2\pi i}{N}\left(m+\frac{1}{2}\right)\left[k+\frac{1}{2}-2^t\left(n+\frac{1}{2}\right)\right] \right],
\end{split}
\label{eq:semiqposrep}
\eeq
and it is convenient to keep this form without performing the geometric sum. For 
\beq
\begin{split}
f_{2SQ}(t)&=\|\tilde{B}_t\|^2=\sum_{k,n=0}^{N/2-1} |\br k |B_t|n \kt|^2\\&=\frac{2^t}{N^2}\sum_{k,n=0}^{N/2-1}\left|\sum_{m=0}^{N/2^t-1} \exp\left[ \frac{2\pi i}{N}\left(m+\frac{1}{2}\right)\left[k+\frac{1}{2}-2^t\left(n+\frac{1}{2}\right)\right] \right] \right|^2=\frac{N}{4},
\end{split}
\label{eq:f2deriv}
\eeq
where the last equality follows on performing the $k$ and $n$ sums first. Note that this also reflects the subunitarity of $\tilde{B}_t$ 
as the norm decreases from $N/2$ at $t=0$ to $N/4$ for all $t>0$.

\begin{figure}[h]
\includegraphics[scale=.62,angle=-90]{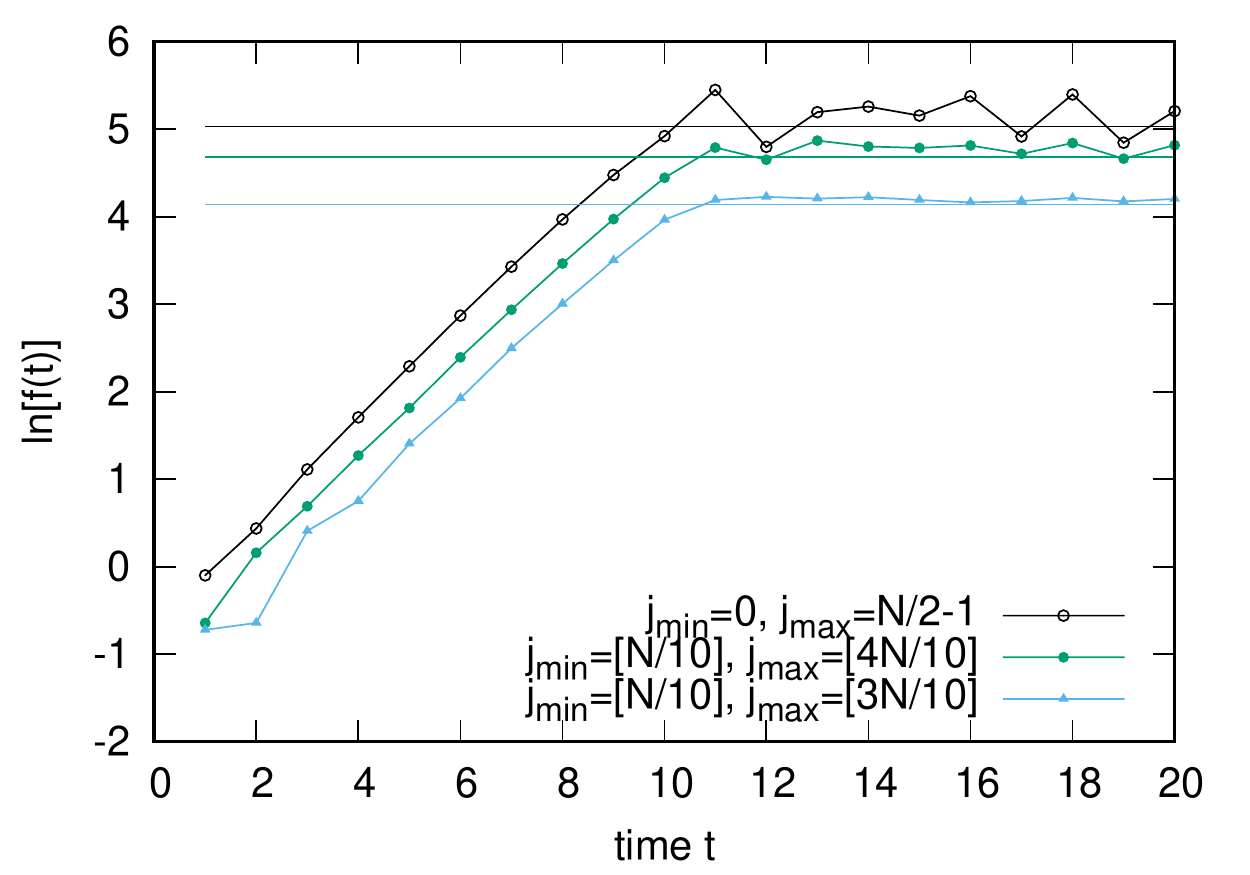}
\includegraphics[scale=.62,angle=-90]{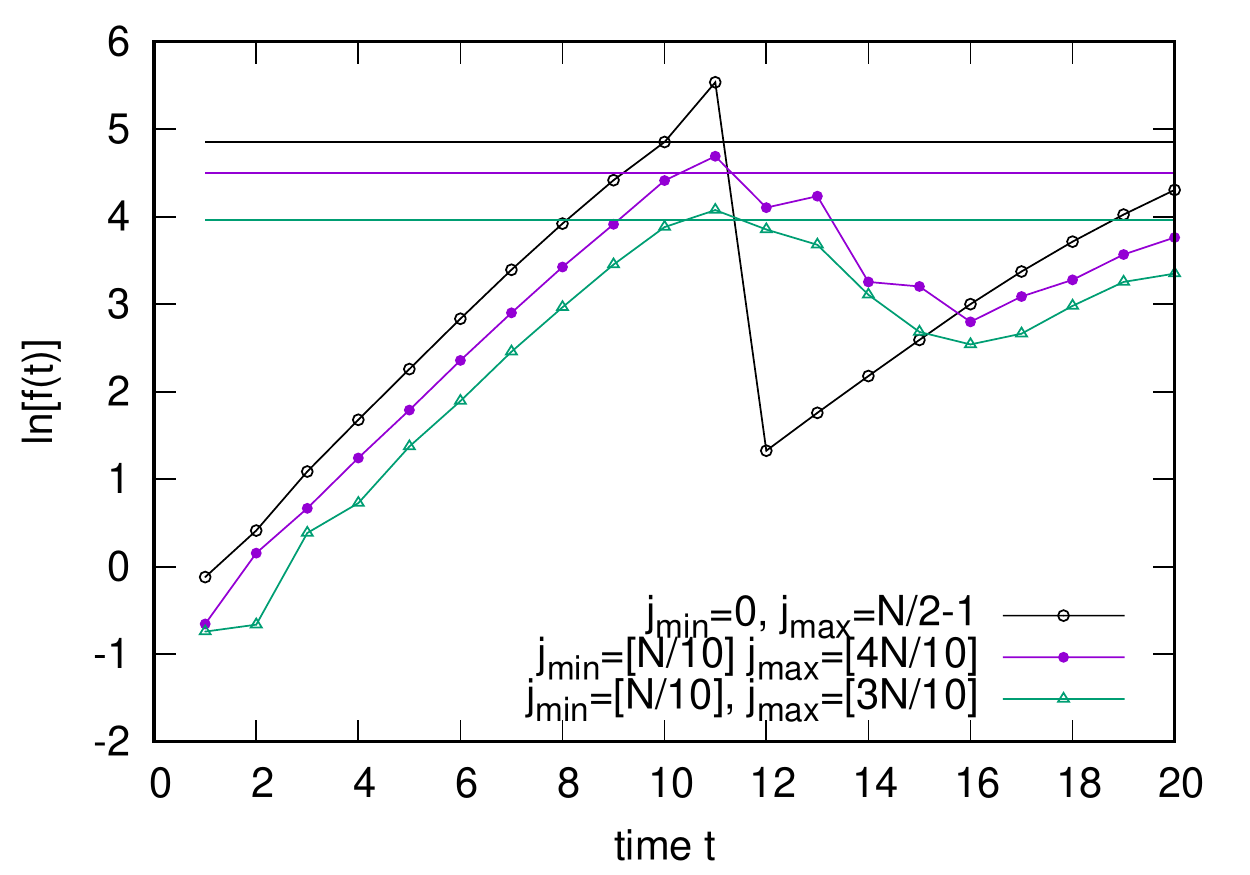}
\caption{ The growth of the commutator's norm $f(t)$ in Eq.~(\ref{eq:foft}) for two values of $N$ and for three projectors. On the left is the case
with $N=2446$ and the right has $N=2048$. The three projectors are as in Eq.~(\ref{eq:proj}) with the $\jmin$ and $\jmax$ values
indicated in the figure. The horizontal lines are from the random matrix saturation value in Eq.~(\ref{eq:avgfoft}).}
\label{fig:bakerotoc1}
\end{figure}
We now turn to the central 	quantity $f(t)$ which measures the noncommutativity of $P(0)$ and $P(t)$ as in Eq.~(\ref{eq:Def_nof_F}).
Figure~(\ref{fig:bakerotoc1}) shows the growth of $f(t)$ for two different values of $N$ and three different position space projectors $P(0)$, one is the $L$ partition that includes the origin which is a fixed point in the classical limit $j_{min}=0$ and $j_{max}=N/2-1$,  one that excludes the origin but is still in the $L$ partition with $j_{min}=[N/10], \; j_{max}=[4 N/10]$, and a third one that is in $L$ but does not include either the origin which is a fixed point or the period-2 orbit at $(1/3,2/3)$. It is observed that the choice of the projector does not make a difference to the growth which is close to being exponential, but the saturation that depends on the size $J$ of the projector and has lesser fluctuations when the partition excludes low-order periodic orbits. The choice of $N=2446$, which is such that $N/2$ is a prime number, ensures that we are far from nongeneric features, while with $N=2048=2^{11}$ an extreme nongeneric case is shown. It is noted that the different partitions have now a more dramatic effect on the $f(t)$, but in fact 
before the log-time the two cases of $N$ and the $3$ partition choices display differences too small to be seen in the figure. It is this growth phase that is also accessible via the semiquantum propagator $B_t$ in Eq.~(\ref{eq:semiqBt}). 

We turn to an analytical derivation that is based on $B_t$ and find the semiquantum approximation of $f(t)$ as
\beq
f_{SQ}(t)=\|\tilde{B}_t\|^2-\|\tilde{B}^{\dagger}_t \tilde{B}_t \|^2.
\label{eq:Def_nof_FSQ}
\eeq
Using the position representation of $B_t$ as given in Eq.~(\ref{eq:semiqposrep}), and considering the projector with 
$\jmin=0$ and $\jmax=N/2-1$ for simplicity, results on further simplifications in 
\beq
\begin{split}
f_{SQ}(t)&= \frac{1}{N^2} \sum_{k=0}^{N/2-1} \sum_{\ovk=N/2}^{N-1} 
\left| \sum_{\ovnu=0}^{2^{t-1}-1}\exp\left[\frac{2 \pi i}{2^t} \nu (k-\ovk)\right] \right|^2 
 \left| \sum_{m=0}^{N/2^t-1}\exp\left[\frac{2 \pi i}{N} m (k-\ovk)\right] \right|^2.
\end{split}
\eeq
Note that the third sum is over $\ovnu$ but the argument contains the complementary $\nu$. 
Further simplifications are indeed possible.
First we have
\beq
\left| \sum_{m=0}^{N/2^t-1}\exp\left[\frac{2 \pi i}{N} m l\right] \right|^2 = \dfrac{\sin^2\left(\pi l/2^t\right)}{\sin^2\left(\pi l/N\right)}. 
\eeq
Then we notice that the $\ovnu<2^{t-1}$ condition implies that the most significant bit of the momentum is $0$, that is the 
periodic orbit, corresponding to the block $\nu$ is below $p=1/2$. Therefore the string $\nu$ has the least significant bit to be $0$, the rest being 
arbitrary. This just implies that $\nu$ is any even integer from $\{0,2, \cdots, 2^t-2\}$, say this is $2n$. 
Then
\beq
\sum_{\ovnu=0}^{2^{t-1}-1}\exp\left[\frac{2 \pi i}{2^t} \nu (k-\ovk)\right]=\sum_{n=0}^{2^{t-1}-1}\exp\left[ 2 \pi i n (k-\ovk)/2^{t-1}\right]=2^{t-1} \delta[ (\ovk-k)\equiv\, 0 \, \text{mod}\, 2^{t-1}],
\eeq
and the semiquantum OTOC reduces to 
\beq
\label{eq:doublesum}
f_{SQ}(t)=\dfrac{2^{2t -2}}{N^2} \sum_{k=0}^{N/2-1} \sum_{\ovk=N/2}^{N-1} \delta[ (\ovk-k)\equiv\, 0 \, \text{mod}\, 2^{t-1}] \dfrac{\sin^2\left(\pi (k-\ovk)/2^t\right)}{\sin^2\left(\pi (k-\ovk)/N\right)}.
\eeq
Due to the Kronecker delta and the numerator in the $\sin^2$ terms only those pairs of $(k,\ovk)$ will contribute whose difference $(\ovk-k)$ is an {\it odd multiple of $2^{t-1}$}.

Let $N=2^T N_0$ where $N_0$ is an odd number $\ge 1$ and $T\ge 1$. The semiquantum propagator $B_t$ is strictly defined for times $ t \le T$. The double sum in Eq.~(\ref{eq:doublesum}) can be reduced further as the argument depends only on the difference $l=(\ovk-k)$ which can take values in $[1,N-1]$.  Let the number of $(\ovk,k)$ pairs that give the same $l$ value be $d_l$. Then 
\beq
d_l=\left \{  \begin{array}{ll} l & 1\le l \le N/2\\N-l & N/2 < l \le N-1.\end{array} \right.
\eeq
Therefore
\beq
f_{SQ}(t)=\dfrac{2^{2t -2}}{N^2} \sum_{l=1}^{N-1} d_l \, \delta[ l \equiv\, 0 \, \text{mod}\, 2^{t-1}] \,\dfrac{\sin^2\left(\pi l/2^t\right)}{\sin^2\left(\pi l/N\right)}.
\eeq
As both $d_l$ and the other quantities being summed over, share
the symmetry $l \rightarrow N-l$, it follows that 
\beq
f_{SQ}(t)=\dfrac{2^{2t -2}}{N^2}\left[ 2\, \sum_{l=1}^{N/2-1} \, l\, 
 \delta[ l \equiv\, 0 \, \text{mod}\, 2^{t-1}] \,\dfrac{\sin^2\left(\pi l/2^t\right)}{\sin^2\left(\pi l/N\right)}+\frac{N}{2} \sin^2(\pi N_0 2^{T-t-1}) \right].
\eeq
Thus whenever $1 \le t \le T-1$ the last term within the bracket, corresponding to $l=N/2$, vanishes. 

Assuming that this is the case, we get on setting $l= (2k+1) 2^{t-1}$
the restriction $k=0,1,\cdots, 2^{T-t-1}-1$, and hence for $1\le t \le T-1$
\beq
f_{SQ}(t)=\frac{2^{t}}{16 \, M^2} \sum_{k=0}^{M-1}\dfrac{2k+1}{\sin^2\left[ \dfrac{\pi (2k+1)}{4 M}\right]}
\label{eq:exactsemiq}
\eeq
with $M=2^{T-t-1}\, N_0=N/2^{t+1}$.
\begin{figure}[h]
\includegraphics[scale=1,angle=-90]{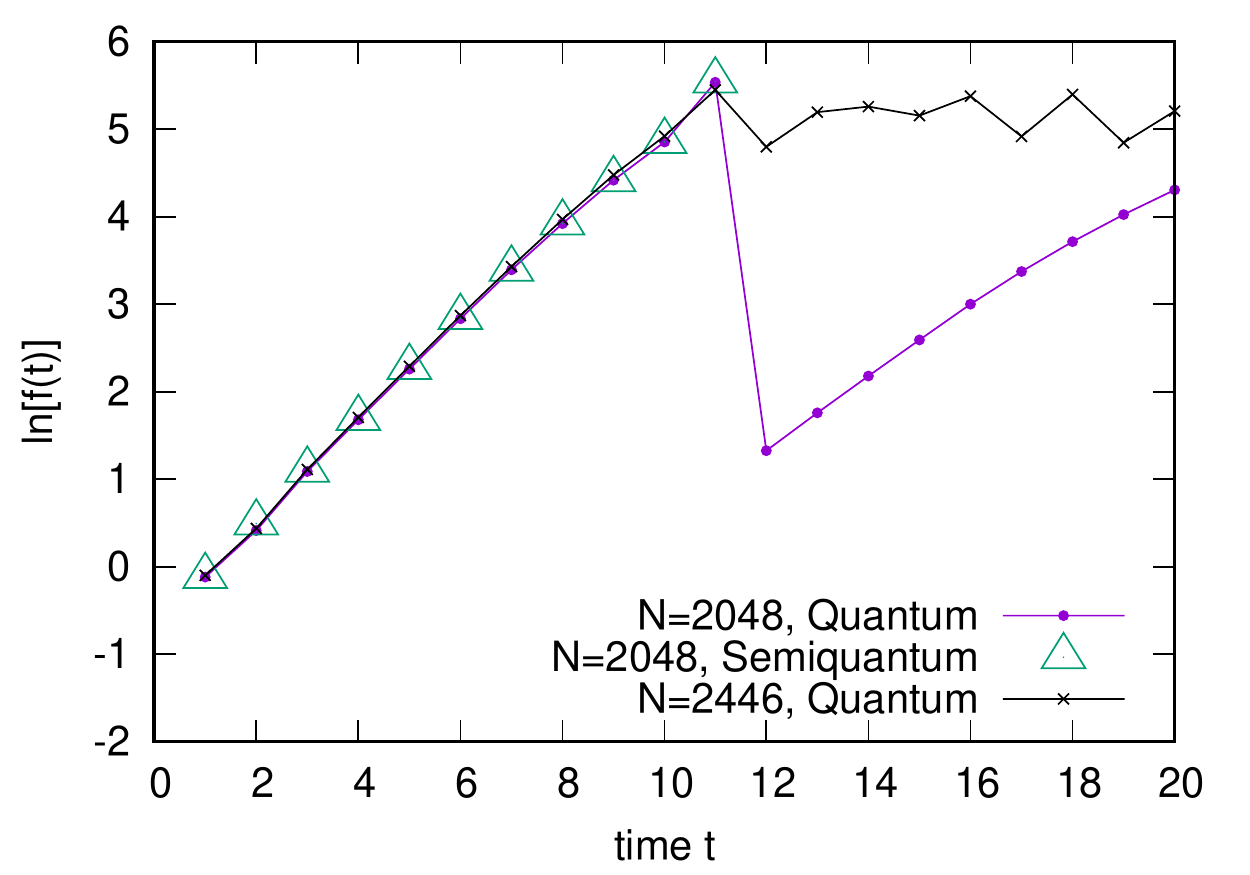}
\caption{ Comparison of the semiquantum analytical evaluation in Eq.~(\ref{eq:exactsemiq}) with the quantum growth of $f(t)$.}
\label{fig:semivsbak}
\end{figure}
In fact for 
$N$ powers of $2$, $N_0=1$ and the following are easily seen to be true:
\beq
f_{SQ}(t=T-2)=\left(2-\frac{1}{\sqrt{2}}\right) 2^{T-5}, f_{SQ}(t=T-1)=2^{T-4}, f_{SQ}(t=T)=2^{T-3}.
\label{eq:fsqT}
\eeq
Figure~(\ref{fig:semivsbak}) compares this semiquantum evaluation with the quantum one for $f(t)$ when $N=1024$. It works well enough 
that visible differences are very small. It is also seen to work well for generic dimensions such as $N=2446$, where $M=[N/2^{t+1}]$
is used in Eq.~(\ref{eq:exactsemiq}). We may conclude that the Lyapunov exponent based on the OTOC is $\ln 2$, except that there is a weak
dependence of time in the coefficient of $2^t$ in Eq.~(\ref{eq:exactsemiq}). To evaluate the coefficient, the sum has to be performed, and for large $M$, it may be replaced by an integral, but this diverges and the singularity at $k=0$ must be compensated by adding and subtracting an appropriate sum to remove the singularity from the integral. 
\beq
\begin{split}
\frac{1}{M^2}\sum_{k=0}^{M-1}\dfrac{2k+1}{\sin^2\left[ \dfrac{\pi (2k+1)}{4 M}\right]} &\approx \frac{1}{2} \int_0^2 \dfrac{x \, dx}{\sin^2(\pi x/4)} -8 \int_0^2 \frac{x\, dx}{\pi^2 x^2} + \frac{16}{\pi^2} \sum_{k=0}^{M-1}\frac{1}{2k+1}\\
&=\frac{8}{\pi^2}\left[1+\ln(8/\pi)+\gamma+\psi_0(M+1/2)\right]\\
&=\frac{8}{\pi^2}\left[1+\ln(8/\pi)+\gamma+\ln M +\mathcal{O}(1/M^2) \right],
\end{split}
\eeq
where $\psi_0(x)$ is the digamma function and $\gamma$ the Euler constant.
Finally then an approximate form of $f_{SQ}(t)$ is
\beq
f_{SQ}(t) =\frac{2^t}{2 \pi^2}\left[ \ln \left( \frac{4\, e^{\gamma+1}}{\pi} \frac{N}{2^t}\right)+\mathcal{O}(2^{t+1}/N)^2\right]. 
\label{eq:approxfsq}
\eeq
This is valid when  $N=N_02^T$, $1\le t \le T-1$ and $M=N/2^{t+1} \gg 1$, that is for times much smaller than the log-time. In practice it appears to be
a good approximation almost close to the log-time. The $\hbar \rightarrow 0$ (here $N \rightarrow \infty$ and $t \rightarrow \infty$) limits cannot be interchanged. Based on the above expression, the following holds:
\beq
\lim_{t\rightarrow \infty}\frac{1}{t} \ln f_{SQ}(t) = \ln 2, 
\eeq
if $N=N_02^{t+t_0}$, where $t_0>1$ and $N_0 \ge 1$ are constants. That is in this limit, the long time limit is slaved to $\hbar \rightarrow 0$. If we want the time $t$ to be fixed, the following is true for instantaneous rates:
\beq
\frac{f_{SQ}(t+1)}{f_{SQ}(t)}\approx 2 \left[ 1-\dfrac{\ln 2}{\ln (N_0 2^{T-t-1})}\right] \rightarrow 2 
\eeq
as $N_0\rightarrow \infty$. Thus the rate also approaches $2$ classically, but very (logarithmically) slowly.
The above expressions are consistent then with 
\beq
f(t) \approx f_{SQ}(t) \sim C_1e^{\lambda t} \, \ln \left(\dfrac{C_2}{e^{\lambda t} \hbar}\right), \;\; t<t_{EF}=\ln(1/\hbar)/\lambda,
\eeq
where $\lambda=\ln 2$ is the classical Lyapunov exponent of the baker map, and $C_1,\,C_2$ are positive constants.

\subsection{Saturation value}

Beyond the log-time, $f(t)$ or the OTOC, saturates and in the bakers map it appears that there is no real gap between the two. The saturation value follows 
if we assume that $U^t$ is chosen from a random set of uniformly distributed unitary matrices of dimension $N$, namely the standard circular unitary ensemble (CUE) of random matrix theory (RMT). While a general result in terms of any operator can be given, we focus on the projection operator as in Eq.~(\ref{eq:proj}) treated above and  use the first equality of Eq.~(\ref{eq:foft}) to write
\beq
f(t)=\sum_{j,j',j''\, \in J} \sum_{\overline{j}\in \overline{J}} U_{jj'}U^*_{\overline{j}j'} U_{\overline{j}j''}U^*_{j j''},
\eeq
where we simply write $U$ for $U^t$ and now treat $U$ as a member of the CUE and 
average over the ensemble. Using
\beq
\begin{split}
\br U_{i_1j_1} U_{i_2j_2} U^*_{i_1'j_1'} U^*_{i_2'j_2'}\kt_{CUE}&=\frac{1}{N^2-1}\left(\delta_{i_1i_1'}  \delta_{i_2i_2'}  \delta_{j_1,j_1'}  \delta_{j_2j_2'} 
+\delta_{i_1i_2'}    \delta_{i_2,i_1'} \delta_{j_1j_2'} \delta_{j_2j_1'} 
\right)\\
&-\frac{1}{N(N^2-1)}\left(\delta_{i_1i_1'}  \delta_{i_2i_2'}  \delta_{j_1,j_2'}  \delta_{j_2j_1'} 
+\delta_{i_1i_2'} \delta_{i_2,i_1'}  \delta_{j_1j_1'}   \delta_{j_2j_2'} 
\right),
\end{split}
\eeq
we get 
\beq
\br f(t)\kt _{CUE}=\frac{J^2(N-J)^2}{N(N^2-1)}.
\label{eq:avgfoft}
\eeq
For the case when $J=N/2$ we get $ \br f(t)\kt _{CUE}=N/16$ for large $N$. It is remarkable that the semiquantum evaluation in Eq.~(\ref{eq:fsqT}) gives $f(T-1)=N/16$ as well and hence when $N$ is a power of $2$, the semiquantum OTOC matches exactly the RMT value at which saturation occurs. The value at $t=T$ is anomalously higher, a feature that seems to hold for all values of $N$. Figure.~(\ref{fig:bakerotoc1}) shows the RMT value for different partitions sizes $J$ and one sees reasonable agreement. Thus for the bakers map, the semiquantum approximation along with the RMT saturation 
value, gives a complete picture of the OTOC or the commutator growth $f(t)$.

\section{Discussions}
The quantization of the bakers map presents an almost exactly solvable model of quantum chaos as far as the OTOC is concerned.
The semiquantum approximation is crucial in making this possible and indicates that along with the exponential growth 
there is also an additional linear time dependence that grows into prominence at the log-time. While the explicit analytical 
evaluation in Eq.~(\ref{eq:exactsemiq}) is for partitions that include the origin, the additional time-dependence
also persists if it does not. It is clear that there cannot be a pure exponential growth as it has to give way to a 
post-log-time growth that  eventually saturates to the RMT value. Whether the form in say Eq.~(\ref{eq:approxfsq}) 
can be generic remains to be seen. Due to the discontinuities in the bakers map, it is known to have anomalous features such as 
additional $\ln N$ terms in semiclassical trace formulas \cite{SaracenoVoros1994,ArulUnusual}. It is possible that these also contribute anomalously to the OTOC. Yet,
the exact nature of the time dependence of the OTOC including an exponential behavior and saturation to a RMT value are generic features.
The $N$ values that are powers of $2$ are also special and have multifractal states and non-generic features, however these only 
dominate the post-log-time phase, when instead of saturating they display oscillations. 
\begin{figure}[h]
\includegraphics[scale=.3]{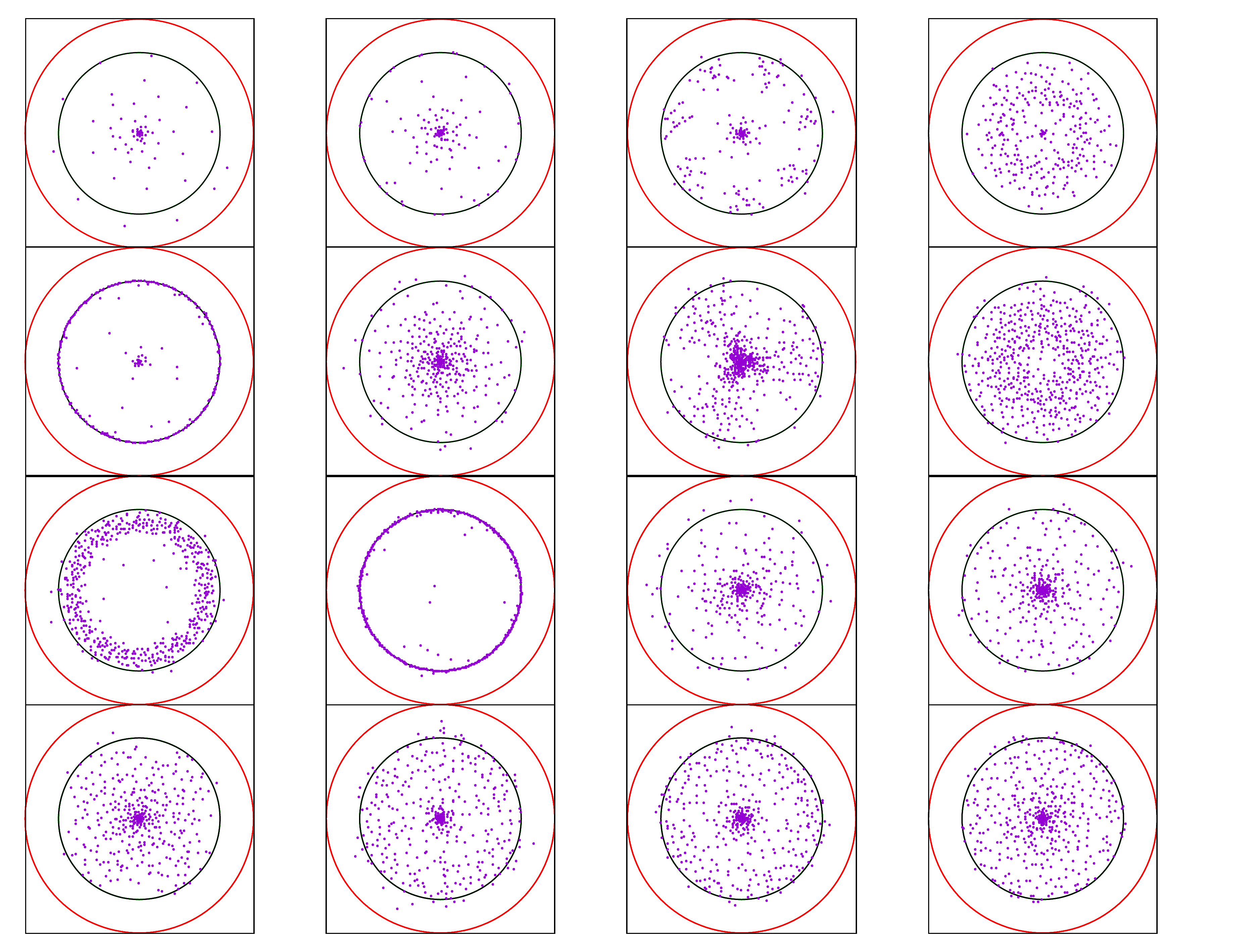}
\caption{The eigenvalues of the $N/2 \times N/2$ left-top corner truncation of $B^t$, that is $\tilde{B^t}$, for $t=1$ to $16$ and $N=1024$. The time increases from left to right and top to bottom. shown are the real and imaginary parts of the eigenvalues, as well as the unit circle and the circle with radius $1/\sqrt{2}$ are shown. }
\label{fig:eigvalsBttrunc}
\end{figure}

Finally as noted in this paper, the truncations of powers of the quantum map determined the OTOC and in fact have 
more information in them. While the OTOC is determined by the singular values, it is also of interest to study the
eigenvalues. Figure.~(\ref{fig:eigvalsBttrunc}) shows the eigenvalues for the first $16$ times when $N=1024$ for the 
truncation with $\jmin=0$ and $\jmax=N/2-1$. It is seen that to begin with a majority of the eigenvalues are very small, 
and that within the log-time period they increase and predominantly occupy the area within the circle of radius $1/\sqrt{2}$. 
Random unitary matrices with such truncations are known to have eigenvalues whose modulus is less that $1/\sqrt{2}$ \cite{Petz2005} and hence
this reflects the way in which the powers of the bakers map randomizes. It is a peculiarity of the bakers map  at $N$ powers of $2$
that the eigenvalues have curious structures at specific times. Notably at $t=\log_2N$ it lies almost wholly on the circle with 
radius $1/\sqrt{2}$ and subsequently again ``collapses" with many eigenvalues being small once more. This is consistent with the 
behavior of $f(t)$ just past the log-time for such powers of $2$ dimensionality.

The operators used in this paper are projectors and hence they have no smooth classical limit. We have verified that using 
operators such as $\cos(2\pi q)$ also lead to the same growth rate in $f(t)$, namely $\approx 2^t$. Thus the rate being 
the classical Lyapunov exponent and not twice it is a consequence of the dynamics itself being diffractive and not 
because of the choice of the observable. Work with smooth maps such as the standard map shows that the rate is approximately 
twice the classical Lyapunov exponent when operators with smooth classical limits are employed. It is hoped that the simple
quantum bakers map, despite its eccentricities, has sufficient genericity that the analytical expressions derived here maybe
of broader applicability or will serve as a foil for less abstract models.

Recent work \cite{GarciaMata2018} has also highlighted an exact evaluation of the OTOC in the quantum cat map and this grows
as $e^{2 \lambda t}$ just as in the standard map \cite{Rozenbaum17}. Additionally the regime beyond the log-time was studied
from the point of view of Ruelle-Pollicott resonances. The present work indicates that the growth of the norm of truncated 
Perron-Frobenius operators, via its singular values, may well reflect the quantum OTOC growth. It maybe noted that the spectrum
of such truncated operators have already been studied in the literature \cite{TruncUnitary5}, but that our discussion above provides  
an impetus for exploring more closely the role of the Perron-Frobenius operator in the growth of the OTOC. While this paper did
not discuss operator scrambling, which does not occur in the (unperturbed) quantum cat map, the baker map's scrambling ability is also accessible via
the semiquantum operator and is part of future work.

\acknowledgments{I am grateful to S. Ganeshan, E. B. Rozenbaum and V. Galitzki for discussions regarding OTOC at an early stage of this work.}

\bibliography{bakerotoc_references}

\end{document}